\documentstyle[twoside,fleqn,espcrc2,psfig]{article}

\def\Journal#1#2#3#4{{#1} {\bf #2} (#4)#3}

\def\PRL{\em Phys. Rev. Lett.}

\def\APJ{\em Astrophys. J} 

\def\lsim{\raise0.3ex\hbox{$\;<$\kern-0.75em\raise-1.1ex
\hbox{$\sim\;$}}}
\def\gsim{\raise0.3ex\hbox{$\;>$\kern-0.75em\raise-1.1ex
\hbox{$\sim\;$}}}

\newcommand{\AmS}{{\protect\the\textfont2
  A\kern-.1667em\lower.5ex\hbox{M}\kern-.125emS}}

\title{Supernova Neutrino Oscillations
\thanks
{Talk presented by H. Nunokawa at Euresco EuroConference Series 
on Frontiers in Particle Astrophysics and Cosmology, 
San Feliu de Guixols, Spain, September 30-October 5, 2000.
}}
\author{H. Nunokawa\address{
Instituto de F\' {\i}sica Gleb Wataghin, 
     Universidade Estadual de Campinas, UNICAMP \\    
     13083-970 Campinas SP, Brazil
        }%
}
       
\begin{document}

\begin{abstract}
We discuss some possible influence of neutrino oscillations 
on physics of supernova mainly focusing on the observations 
of $\bar{\nu}_e$, and present some analysis of SN1987A 
data in the light of three neutrino mixing scheme. 
\end{abstract}

\maketitle

\section{Introduction} 

Sucessful observation of neutrino burst from the supernova
SN1987A, which is believed to be the collapse driven one, 
in the Large Magellanic Cloud by Kamiokande-II~\cite{KamII} 
and IMB~\cite{IMB} detectors have confirmed the basic picture 
of supernova explosion 
(see Refs.~\cite{Suzuki,Raffelt} for a review). 

Such collapse driven supernova can be a good laboratory 
to test various unknown neutrino properties 
not only mass and flavor mixing but also 
other properties such as magnetic moment, life time
and some other non-standard interactions~\cite{Raffelt}. 

It has been known that neutrino flavor conversion 
by the MSW effect~\cite{MSW} inside supernova
could cause some significant influence on supernova 
physics~\cite{Raffelt,SNMSW,fuller,qian}. 
In this talk we review the possible influence of 
neutrino flavor conversion on supernova physics 
and present some analysis of the SN1987A 
data in light of three flavor mixing scheme of neutrinos,
based on our recent work~\cite{MN00}.

\section{Standard picture of supernova neutrino emission}

Here we summarize the basic understanding of neutrinos 
from supernova (SN)~\cite {Suzuki} and their properties 
inside neutrinosphere~\cite {MWS,SNsimu,Janka}. 

A type-II  supernova occurs when a massive star 
($M \gsim 8M_\odot$) has reached the last stage of 
its life. 
Almost all ($\sim 99 \%$) of the gravitational binding energy of 
the final neutron star (about $\sim 10^{53}$ erg) is radiated away in 
form of all flavors of neutrinos. 
The individual total neutrino luminosities in supernovae 
are approximately the same for all flavors but the individual 
neutrino energy distributions are very different\cite {MWS,B87}.
This is because different flavor of neutrinos 
interact differently with the star material, 
as implied by the following reactions, 
\begin{eqnarray}
\label{nu-n}
\nu_e+n&\to & p+e^-,\\
\label{nu-p}
\bar\nu_e+p& \to & n +e^+,\\
\label{nu-N}
\nu +  N & \to &\nu +  N, \ \ (N=p,n).
\end{eqnarray}

Since the cross sections of the charged-current reaction 
is larger than that of the neutral-current one and 
there are more neutrons than protons, 
the $\nu_e$'s have the largest interaction rates with 
the matter and hence thermally decouple at the lowest 
temperature. 
On the other hand, 
$\nu_{\tau(\mu)}$ and $\bar\nu_{\tau(\mu)}$'s lack 
the the charged-current absorption reactions 
on the free nucleons 
inside the neutron star and hence thermally decouple at the 
highest temperature. 

As a result, the average neutrino energies satisfy 
the following hierarchy:
\begin{equation}
\label{hierarchy}
\langle E_{\nu_e}\rangle <\langle E_{\bar\nu_e}\rangle <
\langle E_{\nu_{\tau(\mu)}}\rangle 
\approx\langle E_{\bar\nu_{\tau(\mu)}}\rangle.
\end{equation}
Typically, the average supernova neutrino energies are, 
$\langle E_{\nu_e}\rangle \approx 11-12\ \mbox{MeV},\ 
\langle E_{\bar\nu_e}\rangle
\approx 14-17\ \mbox{MeV},\ \langle E_{\nu_{\tau(\mu)}}\rangle
\approx \langle
E_{\bar\nu_{\tau(\mu)}}\rangle\approx 24-27\ \mbox{MeV}$. 
Such energy hierarchy is of crucial importance for our 
discussion as we will see later.

The shape of the energy spectra of various flavors of neutrinos 
can be described by a "pinched" Fermi-Dirac distribution 
\cite {JH89}. The pinched form can be parametrized by 
introducing an effective "chemical potential", $\eta$ as 
follows, 
\begin{equation}
F(E_\nu) \propto \frac{{E^2}_\nu}{1 + \exp[{-E_\nu/T-\eta}]},
\end{equation}
where $T$ is the temperature. 

There is no physical, significant distinction between 
$\nu_{\mu}$ and $\nu_{\tau}$ 
and their antiparticles in neutrinosphere. 
It is because $\nu_{\mu}$ and ${\bar{\nu}}_{\mu}$ 
are not energetic enough to produce muons by the charged 
current interactions, and the neutral current cross sections of $\nu$ 
and $\bar{\nu}$ are similar in magnitude. 
Therefore, following Ref.~\cite{MN00} we collectively 
denote them as "heavy falvor neutrinos", $\nu_{heavy}$,
in this talk even if they are not heavy (as 
in the case of nverted mass hierarchy we will discuss later).

\section{Possible influence of neutrino oscillation on supernova physics}

Here we briefly review possible effects of neutrino 
oscillation on supernova physics, which were discussed in 
number of previous works. 

First let us stress some 
characteristic features of neutrino oscillations in 
supernova. 
Because of extremely high matter density around neutrinosphere, 
%in the core of the supernova, 
MSW resonant neutrino conversion~\cite{MSW} can occur for the mass 
squared difference much larger than the ones relevant
for solar or atmospheric neutrinos. 
Because of this, in general, one must deal with 
multiple resonances in supernova (in contrast to the
case of solar neutrinos) in the case of 
three (or more) flavor mixing scheme. 

Another important point is that because supernova 
emit all flavors of neutrinos, 
if $\nu_\alpha \to \nu_\beta$ or 
$\bar{\nu}_\alpha \to \bar{\nu}_\beta$ ($\alpha, \beta = e, heavy$)
occurs, $\nu_\beta \to \nu_\alpha$ or 
$\bar{\nu}_\beta \to \bar{\nu}_\alpha$ also must occur simultaneously, 
and the net effect in the case of large oscillation probability 
is the exchange of the energy spectrum
of $\nu_\alpha$ and $\nu_\beta$ or 
$\bar{\nu}_\alpha$ and $\bar{\nu}_\beta$.

\subsection{Shock reheating} 
Neutrino conversion could have significant impact on 
shock re-heating in the delayed explosion scenario~\cite{delayed}. 
If the conversion between 
%$\nu_e$ and $\nu_\mu$ or  $\nu_\tau$ 
$\nu_e$ and $\nu_{heavy}$ 
occurs in the region between the neutrinosphere and 
the stalled shock this can help the explosion~\cite{fuller}. 
Due to the conversion the energy spectra of 
$\nu_e$ and $\nu_{heavy}$ 
% $\nu_e$ and $\nu_\mu$ or $\nu_\tau$ 
can be swapped and
hence $\nu_e$ would have larger average energy leading 
to a larger energy deposition by reactions in 
eqs. (\ref{nu-n}) and (\ref{nu-p}) so that the stalled 
shock would be re-energized. 
In order for this effect to be operative, 
$\Delta m^2 \gsim $ 100 eV$^2$ is required~\cite{fuller}.

\subsection{Heavy elements nucleosynthesis} 

In nature, supernova is considered to be one of the most 
promising site to create neutron rich heavy elements~\cite{Woosley}.
To have successful $r$-process the site must be neutron 
rich, i.e. $Y_e < 0.5$ where $Y_e$ is number of 
electron per baryon. The $Y_e$ value is mainly 
determined by the competition between the 
two absorption reactions in eqs. (\ref{nu-n}) and (\ref{nu-p}). 
In the standard supernova model the latter process 
is favoured due to the higher average energy of 
$\bar\nu_e$ which guarantees the neutron richness. 
If the neutrino oscillations in $\nu_e-\nu_{heavy}$ 
channel 
do occur between the 
neutrinosphere and the region relevant for $r$-process
the site can be driven to proton-rich due to the 
reaction (\ref{nu-n}) and therefore, $r$-process could
be prevented~\cite{qian}. 
Using this argument, mixing parameters in the
region $\Delta m^2 \gsim $ few eV$^2$ and 
$\sin^2 2\theta \gsim 10^{-5}-10^{-4}$ can be
excluded~\cite{qian}.  

\subsection{Influence for ${\nu}_e$ induced events} 

If ${\nu}_e - {\nu}_{heavy}$ oscillation occurs, 
some enhancement of forward peaking elastic scattering 
events at high energies which should be observable 
in water Cherenkov detectors~\cite {MN90}, 
as well as
enhanced oxygen-induced events due to 
a steep rise of the cross section at energies 
higher than $\gsim$ 30 MeV~\cite {Haxton}, 
which could be separated from the dominant isotropic 
$\bar{\nu_e}$ absorption events due to a moderate 
backward peaking of the events~\cite {QF94}.

\subsection{$\bar{\nu}_e$ signal} 

Last argument is the effect of neutrino oscillation 
on $\bar{\nu}_e$ signal in the terrestrial detector, 
which will be discussed more in detail in the following 
sections. 
If large oscillation between 
$\bar{\nu}_e$ and $\bar{\nu}_{heavy}$ occurs, 
the $\bar{\nu}_e$ spectrum gets harder and 
can affect significantly the observation
of supernova neutrinos, 
because the cross section for the $\bar{\nu}_e p$ absorbed 
charged current reaction is much larger than that of 
the $\nu_e e^-$ elastic one. 
Applying this argument, 
% in the context of the two neutrino mixing framework, 
vacuum oscillation and large mixing 
angle MSW solutions to the solar neutrino problem are 
disfavored
by the observed SN1987A neutrino data at
Kamiokande~\cite{KamII} and IMB~\cite{IMB} detectors
~\cite{SSB94,JNR96,Cline00}. 
We will discuss in sec. 5 that 
%this argument could be modified 
how this conclusion could be modified or to be interpreted 
in the context of three neutrino mixing scheme. 

\section{Neutrino conversion in three flavor scheme}

We now discuss the neutrino flavor conversion in supernova (SN), 
in the three flavor mixing scheme. 
To be most conservative, here we assume 
neutrino mixing only among 
three active neutrino flavors, i.e., 
among $\nu_e$, $\nu_\mu$, $\nu_\tau$ and 
among their anti-particles, 
and do not consider the conversion into sterile neutrinos. 

Here, we assume oscillation interpretation of the atmospheric 
~\cite {SKatm} 
and solar neutrino~\cite {solar} data and 
do not consider the LSND result~\cite{LSND}.
Some possible implications of the mixing scheme which 
can explain the LSND result have been discussed in 
Refs.\cite{CFQ,PS00,MY00}. 

We note that two $\Delta m^2$ which can explain atmospheric 
as well as solar neutrinos are irrelevant for 
the shock reheating and 
heavy elements nucleosynthesis (see secs. 3.1 and 3.2) and 
therefor, we will discuss the influence only on  
${\nu}_e$ and $\bar{\nu}_e$ signals, mainly focusing 
on the latter. 

The first very important question is if the neutrino mass 
spectrum adopts the normal or inverted mass hierarchies, 
because it determines wheather there is a resonant conversion 
in the antineutrino channel or not. 
These two mass hierarchies are illustrated schematically 
in Fig. 1, in which we assume that smaller mass squared 
difference is in the range relevant to explain the solar 
neutrino data and the larger one for atmospheric neutrino data. 
\begin{figure}[ht]
%\vglue -0.4cm
\hglue 0.4cm
\psfig{file=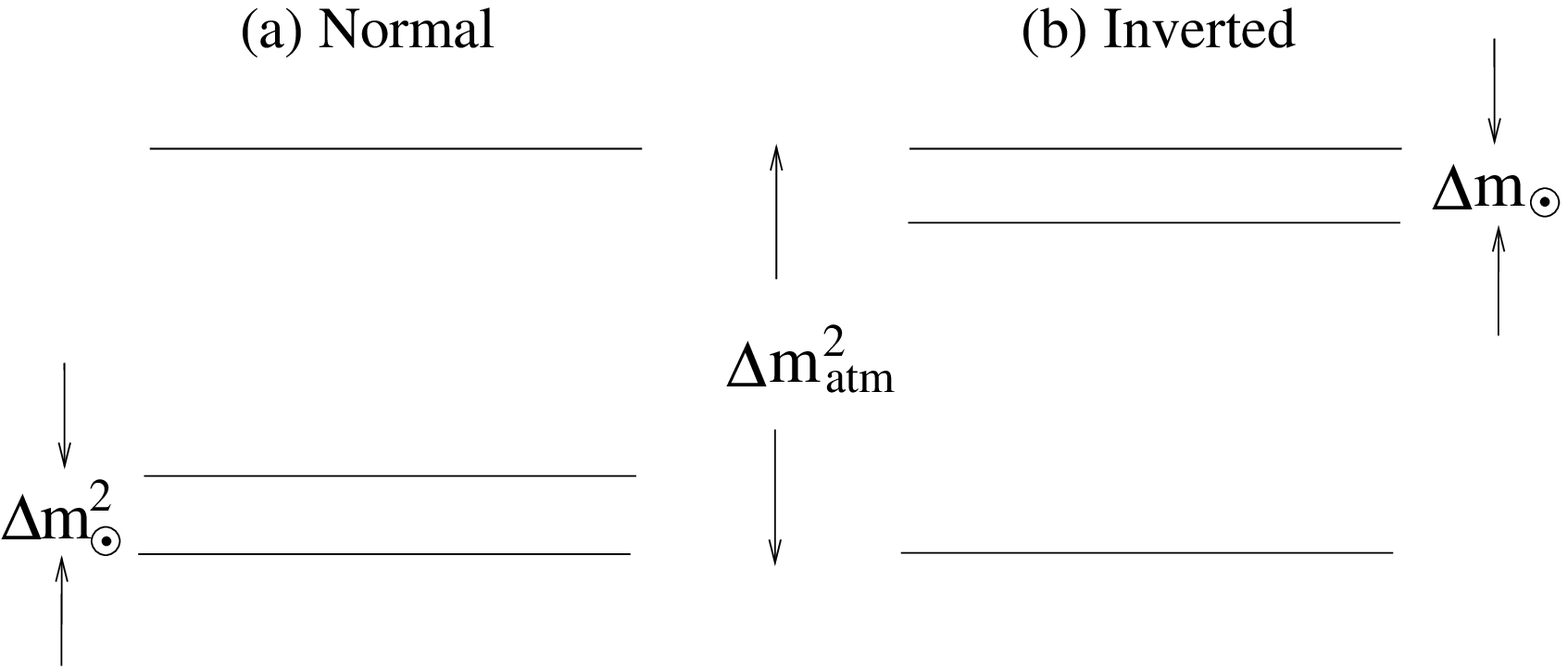,height=4.6cm,width=6.5cm}

\noindent
{\small Fig. 1: Mass hierarchy schemes we consider.} 
\vglue -0.5cm
\end{figure}

\begin{table*}[hbt]
\setlength{\tabcolsep}{1.5pc}
\newlength{\digitwidth} \settowidth{\digitwidth}{\rm 0}
\catcode`?=\active \def?{\kern\digitwidth}
\begin{tabular*}{\textwidth}{@{}l@{\extracolsep{\fill}}rrrr}
\hglue 2.5cm 
\psfig{file=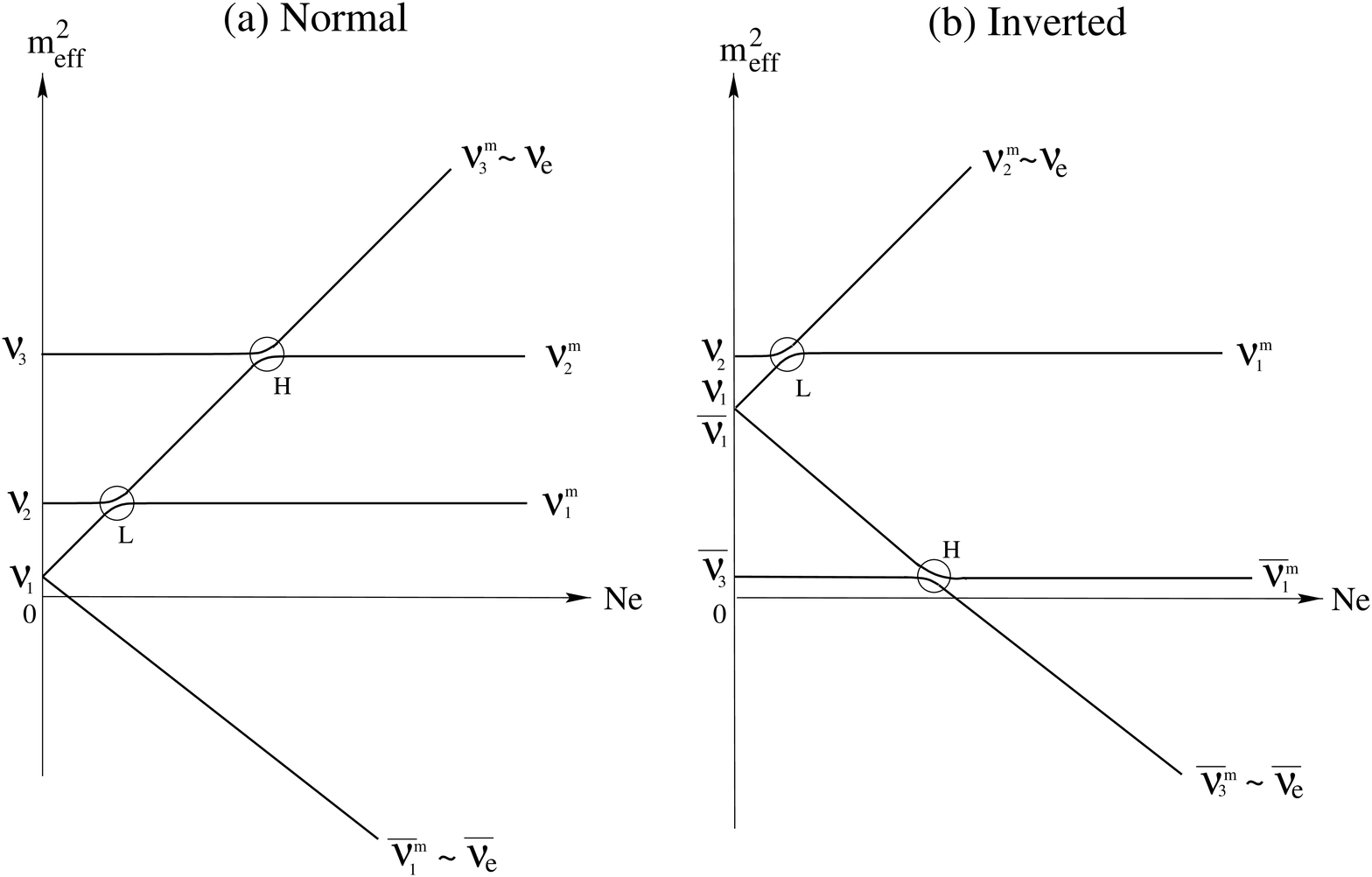,height=6.5cm,width=11.5cm}
\hglue -10.cm 
\noindent \\ 
{\small Fig. 2: 
The schematic level crossing diagram for
the case of (a) normal and (b) inverted mass hierarchies}  
\\
{\small considered in this work. 
The circles with the symbol H and L correspond 
to resonance which occur at higher }\\ {\small 
and lower density, respectively. Adopted from Ref.~\cite{MN00}
}
\end{tabular*}
\vglue -0.5cm 
\end{table*}
In these mass hierarchies, 
the three neutrino and three antineutrino 
eigenstates have two level crossings, 
first at higher (H) density 
and the second at lower (L) density, inside SN as 
schematically illustrated in Fig. 2.
If the mass hierarchies is of normal (inverted) type, 
the H level crossing is in the neutrino (antineutrino) 
channel.

The second important question is that if the neutrino (or antineutrino) 
flavor conversion in SN at H level crossing is adiabatic or not. 
If it is very adiabatic, then the physical properties of neutrino conversion 
is simply $\nu_{e}-\nu_{heavy}$ or 
${\bar{\nu}}_{e}-{\bar{\nu}}_{heavy}$ exchange in the case of 
normal or inverted mass hierarchy. 
It should be emphasized that this feature holds
irrespective of the possible complexity of the solar 
neutrino conversion which governs the L resonance. 
These key features have been pointed out in our 
earlier paper, Ref.~\cite {MN90} in the case of normal
mass hierarchy. 

If the mass hierarchy is inverted and H resonance is
adiabatic, then $\bar{\nu}_e$ spectrum gets harder. 
Since the $\bar{\nu}_e$-induced charged current 
reaction is dominant in water Cherenkov detector, one can severely 
constrain the scenario of inverted mass hierarchy by utilizing 
this feature of neutrino flavor transformation in SN.
When the next supernova event comes it can be used 
to make clear judgement on whether the inverted mass hierarchy is 
realized in nature, a completely independent information from 
those that will be obtained by the future long-baseline neutrino 
oscillation experiments~\cite{JHF,MINOS,OPERA}, 
or by the neutrinoless double beta decay 
experiments~\cite{doublebeta} for the case of Majorana neutrinos. 

The adiabaticity of the H resonance is guaranteed if 
the following adiabaticity parameter $\gamma$ 
is significantly larger than unity at the resonance point:
\begin{eqnarray}
\gamma &&
\equiv \frac{\Delta m^2}{2E}
\frac{\sin^2 2\theta}{\cos 2\theta}
\left|\frac{\mbox{\scriptsize d}\ln N_e}
{\mbox{\scriptsize d}r}\right|^{-1}_{res}\nonumber\\
&& \hskip -0.5cm = 
\left(\frac{\Delta m^2}{2E}\right)^{1-1/n} \hskip -0.5cm 
\frac{\sin^2 2\theta}{(\cos 2\theta)^{1+1/n}}
\ \frac{r_\odot}{n}
\left[\frac{\sqrt{2}G_F\rho_0 Y_e}{m_p}\right]^{1/n}. 
\end{eqnarray}
Here, we assumed that the density profile of 
the relevant region of the star can be described 
as $\rho(r) = \rho_0(r/r_\odot)^{-n}$ 
to obtain the second line in the above equation, 
where $r_\odot = 6.96 \times 10^{10}$ 
cm denotes the solar radius. 
With the choice $n=3$ and 
$\rho_0 \simeq 0.1$ g/cc~\cite{Nomoto}, 
we get, 
\begin{equation}
\gamma \simeq 0.63 \times 
\left[\frac{\sin^2 \theta_{13}}{10^{-4}}\right]
\left[
\frac{\Delta m^2}{10^{-3} \mbox{eV}^2} 
\right]^{2/3}
\left[
\frac{E}{20\ \mbox{MeV}}
\right]^{-2/3}
\hskip -0.1cm ,
\end{equation}
for the small value of $\theta_{13}$. 
Since the conversion probability $P$ is 
approximately given by 
$P_H \simeq \exp[-\frac{\pi}{2}\gamma]$, 
$\sin^2\theta_{13} \gsim \mbox{a few} \times 10^{-4}$
assures adiabaticity in a good accuracy.

For a recent complehensive treatment of neutrino flavor 
conversion in SN in the framework of three-flavor mixing, 
see Ref.~\cite {SD99}.

\section{Analaysis of the supernova SN1987A neutrino 
data in the light of three neutrino scheme
}

While waiting for the next galactic SN, let us perform an analysis 
of the data of neutrinos from SN1987A to gain a hint to the 
problem of the mass and mixing pattern which we want to solve. 

We repeat the similar analyses performed in Refs.~\cite {SSB94,JNR96}
but in the context of three-flavor mixing scheme 
of neutrinos, which is essential for the SN neutrinos. 
In due course, we 
will try to clarify how conclusions obtained in earlier 
works are to be interpreted, or to be conditioned in the 
three-flavor framework. 

We follow the statistical analysis of the SN1987A data 
performed by Jegerlehner, Neubig and Raffelt 
\cite {JNR96} who employed the method of maximum likelihood. 
Using the same Likelihood function which can be found 
in Ref.\cite {JNR96}, 
we perform a fit of the observed data to the 
two parameters, the binding energy of the neutron star, $E_b$, 
and the $\bar{\nu}_e$ temperature, $T_{\bar{\nu}_e}$. 

First we show the result of our analysis 
without assuming neutrino oscillation. 
In Fig. 3 we show the contours of constant likelihood 
for Kamiokande and IMB data, as well as the combined one. 

For simplicity, as in Ref.~\cite {JNR96}, 
we set the ``effective'' chemical potential 
equal to zero in the neutrino distribution 
functions because we believe that our results 
would not depend much even if we introduce
some non-zero chemical potential. 
We note that our result is in very good agreement 
with the one obtained in Ref.~\cite{JNR96}. 

We note that in Fig. 3, the agreement between Kamionde 
and IMB data is not so significant though these data
are consistent. 
The possibility to reconcile the difference between 
Kamionde and IMB data by utilizing the earth matter
effect has been recently considered in Ref.~\cite{LS00}. 
Here, we do not enter into this point and simply 
try to do the combined fit as in Ref.~\cite{JNR96}. 

%
%\vglue -0.5cm 
\hglue -0.8cm 
\psfig{file=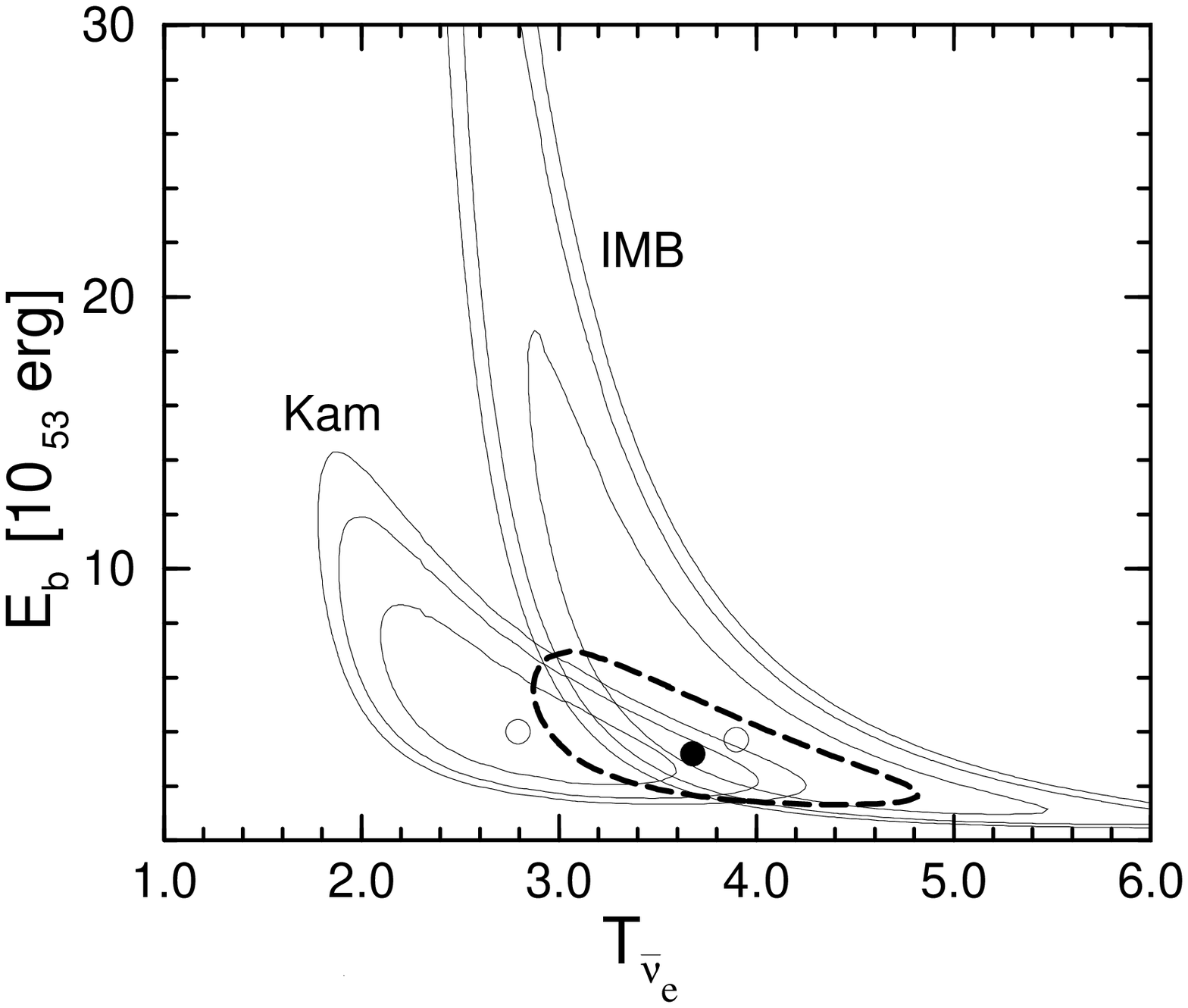,height=7.8cm,width=8.2cm}
\vglue -0.2cm 
\noindent{\small 
Fig. 3: 
Contours of constant likelihood corresponding to 
68.3, 90 and 95.4 \% C.L. region in 
$E_b - T_{\bar{\nu}_e}$  for Kamiokande and IMB data. 
Best fitted points are indicated by open circles. 
Think dashed curve shows the combined 
95.4 \% C.L. contour and the solid circle 
indicates the combined best fit. 
}
\vglue 0.2cm 
\label{Fig3}

\subsection{Case I}

Let us first consider the case where 
the observable effect is most significant: 
the mass hierarchy is inverted and 
H resonance is very adiabatic, which can be realized if 
$\sin^2\theta_{13} \gsim 10^{-4}$. 
In this case there is a efficient conversion in $\bar{\nu}$ 
channel at H resonance.  

We draw in Fig. 4 equal likelihood contours 
 as a function of the heavy to light temperature 
 ratio $\tau$ on the space spanned by 
 $\bar{\nu}_e$ temperature and total neutrino 
 luminosity by giving the neutrino events 
 from SN1987A observed by Kamiokande~\cite {KamII} 
and  IMB~\cite{IMB} detectors.  
We characterize the difference in energy by the temperature 
ratios of $\nu_e$ and $\bar{\nu}_e$ to $\nu_{heavy}$ as 
\begin{equation}
\tau \equiv 
\frac{T_{\nu_{heavy}}}{T_{\bar{\nu}_e}} \simeq
\frac{T_{\bar{\nu}_{heavy}}}{T_{\bar{\nu}_e}}. 
\end{equation}
According to the simulation of supernova dynamics which is 
carried out in Ref.~\cite{MWS,SNsimu,Janka}, 
typically $\tau \simeq 1.4-2.0$.

\vglue -0.1cm 
\hglue -1.0cm 
\psfig{file=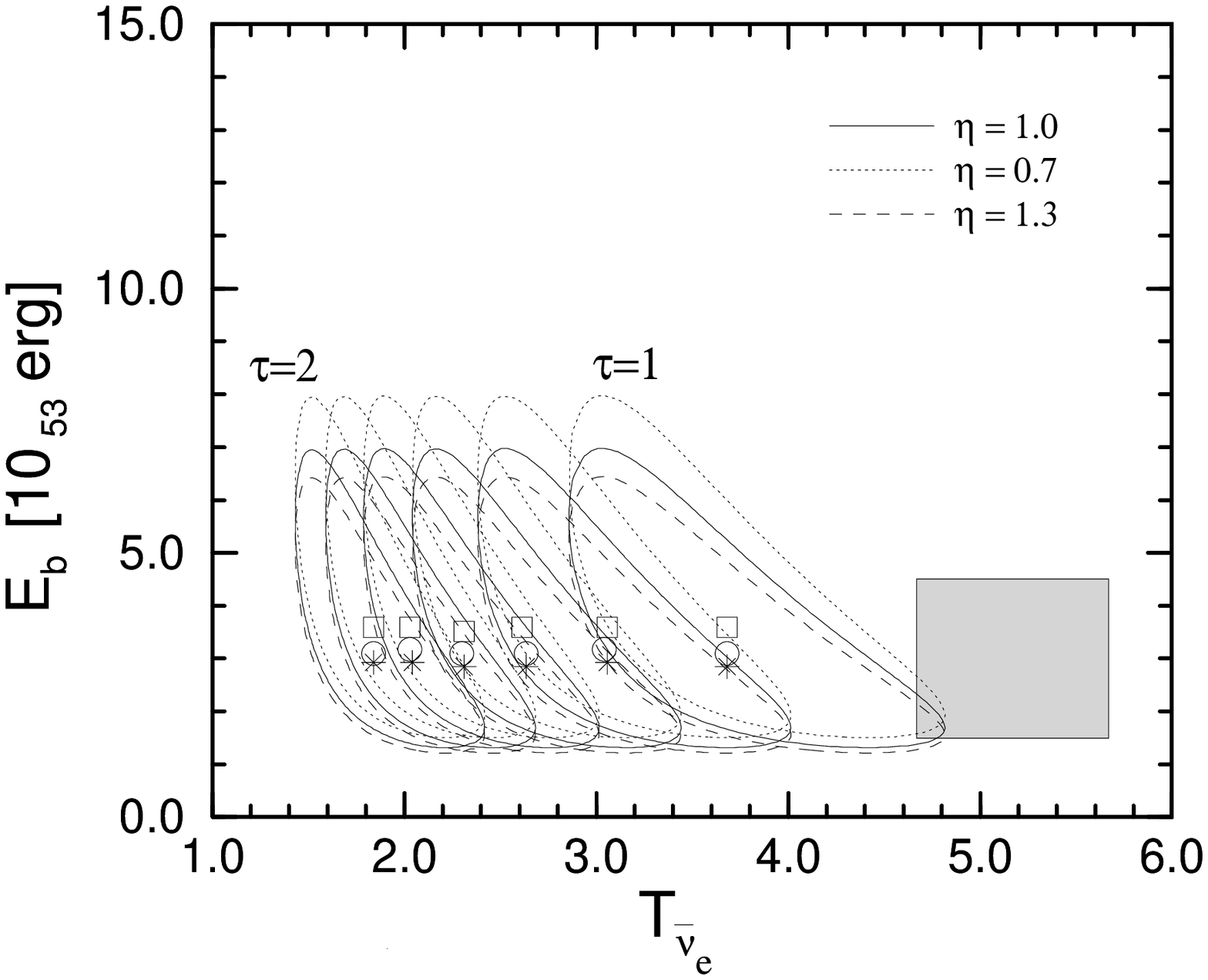,height=8.5cm,width=8.5cm}
%}}
\vglue -0.2cm 
\noindent{\small Fig. 4: 
Contours of constant likelihood which correspond to
95.4 \% confidence regions for the inverted mass hierarchy 
under the assumption of adiabatic H resonance.
{}From left to right,
$\tau \equiv T_{\bar{\nu}_{heavy}}/T_{\bar{\nu}_e} =
T_{{\nu_{heavy}}}/T_{\bar{\nu}_e} =  2, 1.8, 1.6, 1.4, 1.2$ and 1.0. 
Best-fit points for
$T_{\bar{\nu}_e}$ and $E_b$ are also shown
by the open circles.
The parameter $\eta$ parametrizes the departure from the 
equipartition of energy,  
$ L_{\nu_{heavy}} = L_{\bar{\nu}_{heavy}}
= \eta L_{\nu_e} = \eta L_{\bar{\nu}_e}$,
and 
the dotted lines (with best fit indicated by open squares) and
the dashed lines (with best fit indicated by stars) 
are for the cases $\eta = 0.7$ and 1.3, respectively.
Theoretical predictions from supernova models 
are indicated by the shadowed box. 
Adopted from Ref.~\cite{MN00}
}
\vglue 0.5cm

In addition to it we introduce an extra 
parameter $\eta$ defined by 
$L_{\nu_{heavy}}  = L_{\bar{\nu}_{heavy}} 
= \eta L_{\nu_e} = \eta L_{\bar{\nu}_e}$
which describe the departure from equipartition 
of energies to three neutrino species and examine the 
sensitivity of our conclusion against the change in the 
SN neutrino spectrum. 

At $\tau = 1$, that is at equal $\bar{\nu}_e$ 
and $\nu_e$ temperatures, the 95 $\%$ likelihood 
contour marginally overlaps with the theoretical 
expectation~\cite{Janka} represented by the 
shadowed box in Fig. 4.
When the temperature ratio $\tau$ is varied 
from unity to 2 the likelihood contour moves 
to the left, indicating less and less consistency, 
as $\tau$ increases, between the standard 
theoretical expectation and the observed feature 
of the neutrino events after the MSW effect 
in SN is taken into account.
This is simply because the observed energy 
spectrum of $\bar{\nu}_e$ must be interpreted 
as that of the original one of $\bar{\nu}_{heavy}$,
in the presence of the MSW effect in 
the anti-neutrino channel, which implies that 
the original $\bar{\nu}_e$ temperature
must be lower by a factor $\tau$ than 
the observed one, leading to stronger 
inconsistency at larger $\tau$.

The solid lines in Fig. 4 are for the case 
of equipartition of energy into three flavors, 
$\eta = 1$, whereas the dotted and the dashed 
lines are for $\eta = 0.7$ and 1.3, respectively.
We observe that our result is very 
insensitive against the change in $\eta$.

We conclude that if the temperature ratio 
$\tau$ is in the range 1.4-2.0 as the SN 
simulations indicate, the inverted hierarchy 
of neutrino masses is disfavored by the neutrino 
data of SN1987A unless the H resonance 
is nonadiabatic. 
We note that similar conclusion is obtained 
also in Ref.~\cite{LS00}.  

\subsection{Case II}

Let us next consider the case where mass hierarchy is normal
or inverted but H resonance is very non-adiabatic, which
correspond to the case with very small $\theta_{13}$, 
$\sin^2\theta_{13} \ll 10^{-4}$. 
In this case, there is no (or no significant) adiabatic 
conversion in $\bar{\nu}$ channel, and therefore, 
as far as $\bar{\nu}_e$ signal is concerned, 
the problem is essentially reduced to that in the 
two flavor scheme which was studied in Refs.~\cite{SSB94,JNR96}. 

Conclusions can be summarized as follows: 
If the small mixing angle (SMA) MSW is the solution to the solar 
neutrino problem there is only a minor effect because neither 
the vacuum oscillation nor the earth matter effects are effective 
because of small $\theta_{12}$. 
If the large mixing angle (LMA) or 
low $\Delta m^2$ (LOW) MSW or 
vacuum oscillation (VO) is the solution, 
we have a potential trouble because a good 
fraction of $\bar{\nu}_e$ is transformed 
into $\bar{\nu}$$_{heavy}$ and vice versa.

We have repeated the same analysis as done
in Ref.\cite{JNR96} and obtained very similar results. 
We conclude that VO and LOW solutions are more 
disfavored than the LMA MSW solution mainly 
because of the absence of earth matter effect
for the former solutions as discussed 
in Refs.~\cite{SSB94,JNR96}. 

\vglue 0.1cm 
\hglue -0.2cm 
\psfig{file=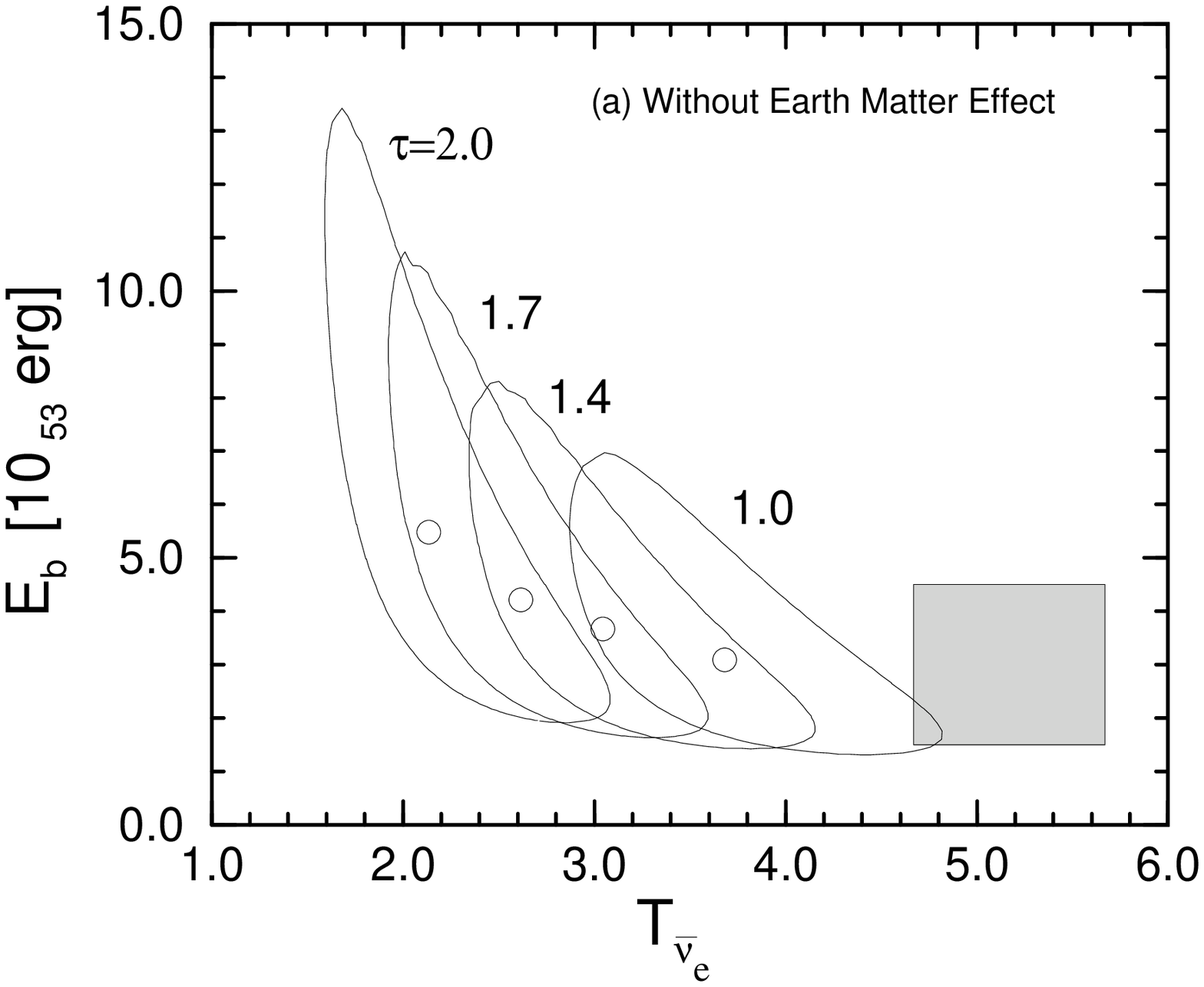,height=6.5cm,width=7.6cm}
\vglue -0.5cm 
\hglue -0.2cm 
\psfig{file=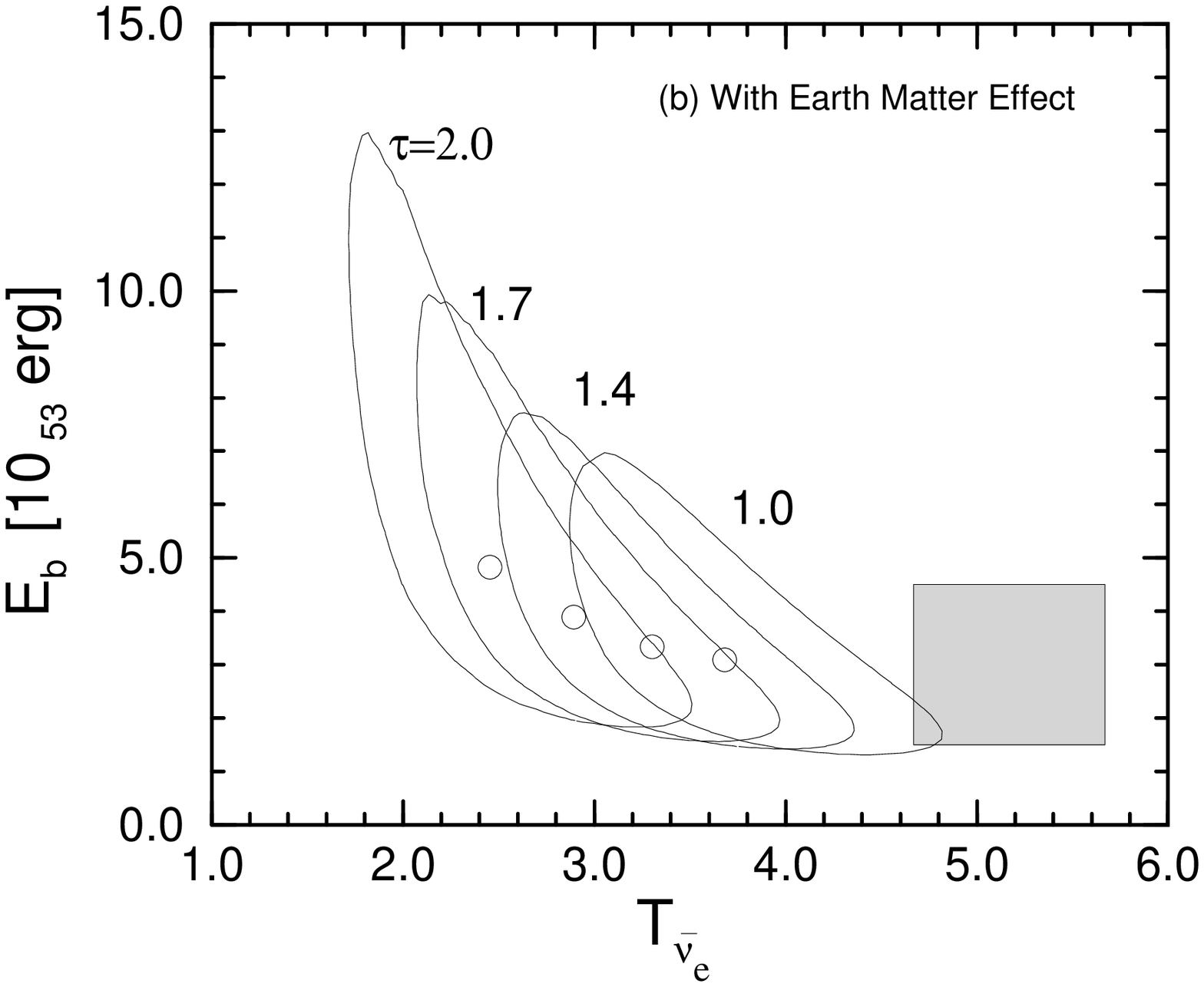,height=6.5cm,width=7.6cm}
\vglue -0.1cm 
\noindent{\small Fig. 5: 
Contours of constant likelihood corresponding to 
95.4 \% C.L. for the large mixing
angle MSW solution (a) without and (b) with 
earth matter effect. We have taken 
mixing parameters as 
$\Delta m^2 = 3\times 10^{-5}$ eV$^2$ and
$\sin^2 2\theta = 0.8$ for LMA MSW solution. 
Adopted from Ref.~\cite{MN00}
}
\label{Fig3}
\vglue 0.3cm 

Let us try to demonstrate explicitly this point. 
We employ a particular set of parameters of 
the LMA MSW solution and compare the behavior 
of the likelihood contours with and without 
earth matter effect. 
We show our results in Fig. 5. 
For simplicity, we set $\theta_{13}$ = 0 but our result does not 
change much as long as the parameter is under the CHOOZ bound 
\cite {CHOOZ}.  
We see that inconsistency between the fitted data
and theoretical prediction is somewhat weakened 
when earth matter effect is included. 

\subsection{Case III}

Finally, let us consider the case where mass hierarchy is inverted 
but H resonance is moderately adiabatic, which means that 
$P_H$ is not so close to zero or to unity, 
which can correspond to the case with 
$\sin^2\theta_{13} \sim 10^{-4} - 10^{-5}$.

In this case, 
$\bar{\nu}_e$-${\bar{\nu}}_{heavy}$ 
transformation occurs 
with the probability $1 - P_H$, and it would imply 
the similar but milder effect than that we have 
obtained with a good adiabaticity of the H resonance, 
shown in Fig. 4. 
If the next galactic supernova is detected 
by Superkamiokande, then we will be able 
to discriminate the moderately nonadiabatic 
case from the adiabatic one. 

\section{Summary}
We have discussed the possible influence of neutrino 
oscillation for supernova physics, in particular 
for observation of $\bar{\nu}_e$ spectrum. 
We stress that the mass spectrum can be tested by 
$\bar{\nu}_e$ signal from supernovae, if 
$\theta_{13}$ is not very small. 
We performed some analysis of SN1987A data in the 
context of three flavor mixing and conclude that
inverted mass hierarchy is disfavored by the
data unless $\theta_{13}$ is very small,  
$ \sin^2 \theta_{13} \lsim 10^{-4}$.  
We hope that future galactic supernova may 
provide more clear information. 

\section*{ACKNOWLEDGMENTS}
The author thanks H. Minakata for collaboration, 
A. Yu Smirnov, C. Lunardini, M.\ C.\ Gonzalez-Garcia, 
and R. Tom\`as for useful discussions. 
This work was supported by the Brazilian funding 
agency Funda\c{c}\~ao de Amparo \`a Pesquisa do 
Estado de S\~ao Paulo (FAPESP). 

%%%%%%%%%%%%%%%%%%%%%%%%%% Bibliography %%%%%%%%%%%%%%%%%%%%%%%%%%%%%%


\begin{thebibliography}{99}

\bibitem {KamII}
K. S. Hirata et al., Phys. Rev. Lett. {\bf 58} (1987) 1490;
Phys. Rev. {\bf D38} (1988) 448. 

\bibitem {IMB}
R. M. Bionta et al., Phys. Rev. Lett. {\bf 58} (1987) 1494. 

\bibitem {Suzuki}
H. Suzuki, in {\it Physics and Astrophysics of Neutrinos}, 
edited by M. Fukugita and A. Suzuk, Springer Verlag, Tokyo, 1994. 


\bibitem{Raffelt}
G. G. Raffelt, 
{\it Stars as Laboratories for Fundamental Physics}, 
Univ. of Chicago Press, Chicago, 1996; 
See also talk by G. Raffelt, in these proceedings. 
 
\bibitem{MSW}
S. P. Mikheyev and A. Yu. Smirnov, Nuovo Cim. {\bf 9C} (1986) 17;
L. Wolfenstein, Phys. Rev. {\bf D17} (1978) 2369.


\bibitem{SNMSW}
L. Wolfenstein, Phys. Rev. {\bf D20} (1979) 2634;
S. P. Mikheyev and A. Yu. Smirnov, Proceedings of 6th Moriond 
Workshop on Massive Neutrinos in Astrophysics and Particle 
Physics, Tignes, France, edited by O. Fackler and J. Tran Thanh Van 
(Editions Frontieres, Gif-sur-Yvette, 1986) page 355. 
%
D. N\"otzold, Phys. Lett. {\bf B196} (1987) 315; 
J. Arafune, M. Fukugita, T. Yanagida, and M. Yoshimura, 
Phys. Rev. Lett. {\bf 59} (1987) 1864;
H. Minakata, H. Nunokawa, K. Shiraishi, and H. Suzuki, 
Mod. Phys. Lett. {\bf A2} (1987) 827;
H. Minakata and H. Nunokawa, Phys. Rev. {\bf D38} (1988) 3605;
S. P. Rosen, ibid. {\bf D37} (1988) 1682;
T. K. Kuo and J. Pantaleone, ibid. {\bf D37} (1988) 298;
T. P. Walker and D. N. Schramm, Phys. Lett. {\bf B195} (1987) 331;
%
% Y.-Z, Qian, G. M. Fuller, G. J. Mathews, R. Meyle, J. R. Wilson, and 
% S. E. Woosley, Phys. Rev. Lett. {\bf 71} (1993) 1965;
G. M. Fuller, J. R. Primack, and Y.-Z, Qian, 
Phys. Rev. {\bf D52} (1995) 1288;
%
A. Barrows, D. Klein, and R. Gandhi, 
ibid. {\bf D45} (1992) 3361;
G. M. Fuller, W. C. Haxton, and G. C. McLaughlin, 
ibid. {\bf D59} (1999) 085005.

\bibitem{fuller}
G. M. Fuller {\it et al.}, 
\Journal{\APJ}{389}{ 517}{1992}.

\bibitem{qian}
Y.-Z. Qian {\it et al.}, 
\Journal{\PRL}{71}{ 1965}{1993}.


\bibitem {MN00}
H. Minakata and H. Nunokawa, hep-ph/00010240. 

\bibitem {MWS}
R. Meyle, J. R. Wilson, and D. N. Schramm, Astrophys. J. 
{\bf 318} (1987) 288.


\bibitem {SNsimu}
J. R. Wilson, R. Meyle, S. Woosley, and T. Weaver, Ann. N.Y. Acad. Sci. 
{\bf 470} (1986) 267;
A. Barrows and J. M. Lattimer, Astrophys. J. 
{\bf 307} (1986) 178.

\bibitem {Janka}
H.-T. Janka, in {\it Vulcano Workshop 1992; Frontier Objects in 
Astrophysics and Particle Physics}, Proceedings of the Workshop 
Vulcano, Italy, 1992, edited by F. Giovannelli and G. Mannochi, 
IPS Conf. Proc. No. 40 (Italian Physical Society, Vulcano, 1993). 

\bibitem {B87}
S. W. Bruenn, Phys. Rev. Lett. {\bf 59} (1987) 938;
A. Barrows, Astrophys. J. {\bf 334} (1988) 891;
E. S. Myra and A. Barrows, ibid. {\bf 364} (1990) 222.

\bibitem {JH89}
H.-T. Janka, Astron. Astrophys. {\bf 224} (1989) 49;
Astron. Astrophys. Suppl. {\bf 78} (1989) 375;
P. M. Giovanoni, P. C. Ellison, and S. W. Bruenn, 
Astrophys. J. {\bf 342} (1989) 416. 



\bibitem{delayed}
J. R. Wilson, {\it Numerical Astrophysics} ed. 
J. M. Centrella, J. M. Leblanc and R. L. Bowers, p.422 
(Boston, Jones and Bartlett, 1983).




\bibitem{Woosley}
S. E. Woosley and E. Baron,  Astrophys. J. {\bf 391} (1992) 228; 
S. E. Woosley,  Astron. Astrophys. Suppl. Ser. {\bf 97} (1993) 205;
S. E. Woosley and R. D. Hoffman,  Astrophys. J. {\bf 395} (1992) 202;
B. S. Meyer {\it et al.,} Astrophys. J.  {\bf 399} (1992) 656; 
S. E. Woosley {\it et al.}, Astrophys. J. {\bf 433} (1994) 229.

\bibitem {MN90}
H. Minakata and H. Nunokawa, Phys. Rev. {\bf D41} (1990) 2976.

 \bibitem {Haxton}
 W. Haxton, Phys. Rev. {\bf D36} (1987) 2283.

 \bibitem {QF94}
 Y.-Z. Qian and G. M. Fuller, Phys. Rev. {\bf D49} (1994) 1762.

\bibitem{SSB94} 
A. Yu. Smirnov, D. N. Spergel, and J.N. Bahcall, 
Phys. Rev. {\bf D49} (1994) 1389.

\bibitem {JNR96} 
B. Jegerlehner, F. Neubig and G. Raffelt, 
Phys. Rev. {\bf D54} (1996) 1194.

\bibitem {Cline00} 
D. Cline, 
% Talk at Europhysics Neutrino Oscillation Workshop, NOW2000, 
% September 9-16, 2000, Otranto, Italy.
astro-ph/0010339. 

\bibitem {SKatm}
Y. Fukuda et al. (Kamiokande collaboration), 
Phys. Lett. {\bf B335} (1994) 237;
Y. Fukuda et al. (SuperKamiokande collaboration), 
Phys. Rev. Lett. {\bf 81} (1998) 1562;
T. Kajita, in {\it Neutrino Physics and Astrophysics}, 
Proceedings of the XVIIIth International Conference on Neutrino 
Physics and Astrophysics (Neutrino '98), June 4-9, 1998, Takayama, 
Japan, edited by Y. Suzuki and Y. Totsuka, 
(Elsevier Science B.V., Amsterdam, 1999) page 123.


\bibitem {solar}
Homestake Collaboration, K. Lande {\it et al.},
Astrophys\ .J.\ {\bf 496}, 505 (1998); 
%
SAGE Collaboration, J.\ N.\ Abdurashitov {\it et al.},
Phys.\ Rev.\ C {\bf 60}, 055801 (1999); 
% 
GALLEX Collaboration, W.\ Hampel {\it et al.}, Phys.\
Lett.\  B {\bf447}, 127 (1999); 
% 
Kamiokande Collaboration, Y. Fukuda  {\it et al.} 
Phys. Rev. Lett. {\bf 77}, 1683 (1996);
%
SuperKamiokande Collaboration,  Y.\ Fukuda {\it et al.}, 
Phys. Rev. Lett. {\bf 81}, 1158 (1998); 
{\it ibid.}  {\bf 81}, 4279 (1998); 
{\it ibid.}  {\bf 82}, 2430 (1999); 
{\it ibid.}  {\bf 82}, 1810 (1999). 

\bibitem{LSND}
C. Athanassopoulos {\it et al. }, for LSND collaboration, 
Phys. Rev. Lett. {\bf 77} (1996) 3082; 
Phys. Rev. C{\bf 54} (1996) 2685. 


\bibitem {CFQ}
D. O. Caldwell, G. M. Fuller and Y-Z. Qian, 
Phys. Rev. {\bf D61} (2000) 123005.  

\bibitem {PS00}
O. L. G. Peres and A. Yu. Smirnov, hep-ph/0011054; 
See also talk by O. L. G. Peres in these proceedings. 
 
\bibitem {MY00}
H. Murayama and T. Yanagida, hep-ph/0010178. 

\bibitem {JHF}
JHF Neutrino Working Group, Y. Itow et al., 
Letter of Intent:
A Long Baseline Neutrino Oscillation Experiment 
using the JHF 50 GeV Proton-Synchrotron 
and the Super-Kamiokande Detector, February 3, 2000, 
http://neutrino.kek.jp/jhfnu

\bibitem {MINOS}
The MINOS Collaboration, P. Adamson et al., 
MINOS Detectors Technical Design Report, Version 1.0, 
NuMI-L-337, October 1998.

\bibitem {OPERA}
OPERA Collaboration, M. Guler et al., 
OPERA: An Appearance Experiment to Search for Nu/Mu 
$\leftarrow$$\rightarrow$ Nu/Tau 
Oscillations in the CNGS Beam. Experimental Proposal, 
CERN-SPSC-2000-028, CERN-SPSC-P-318, LNGS-P25-00, Jul 2000. 


\bibitem {doublebeta}
H. Minakata, Talk at Workshop on Neutrino Oscillations and Their Origin, 
Fujiyoshida, Japan, February 11-13, 2000, 
hep-ph/0004249, to appear in Proceedings;
F. Vissani, JHEP {\bf 9906} (1999) 022; 
H.V.Klapdor-Kleingr\"othaus, H. P\"as and A. Yu. Smirinov, hep-ph/0003219.

\bibitem {Nomoto}
K. Nomoto and M. Hashimoto, Phys. Rep. {\bf 163} (1988) 13. 
See also the figures in H. Minakata and H. Nunokawa in~\cite {SNMSW}.


\bibitem {SD99}
A. S. Dighe and A. Yu. Smirnov, Phys. Rev. {\bf D62} (2000) 033007.

\bibitem {LS00}
C. Lunardini and A. Yu. Smirnov, hep-ph/0009356;
See also the talk by C. Lunardini in these proceedings. 

\bibitem {CHOOZ}
CHOOZ Collaboration, M. Apollonio et al., 
Phys.\ Lett.\ B{\bf 420} (1998) 397; 
{\it ibid.}, B{\bf 466} (1999) 415. See also, 
The Palo Verde Collaboration, F. Boehm et al., hep-ex/0003022.  

\end{thebibliography}
\end{document}